\documentclass[showpacs,twocolumn,prb,aps,superscriptaddress]{revtex4}
\usepackage{amssymb}
\usepackage{amsmath}
\usepackage{graphicx}

\begin{document}

\title{Quantum percolation in quantum spin Hall antidot systems}
\author{Rui-Lin Chu}
\affiliation{Department of Physics, The University of Hong Kong, Pokfulam Road, Hong Kong}
\author{Jie Lu}
\affiliation{College of Physics, Hebei Advanced Thin Films Laboratory, Hebei Normal University, Shijiazhuang,China}
\affiliation{Department of Physics, The University of Hong Kong, Pokfulam Road, Hong Kong}
\author{Shun-Qing Shen}
\affiliation{Department of Physics, The University of Hong Kong, Pokfulam Road, Hong Kong}
\date{\today}

\begin{abstract}
We study the influences of antidot-induced bound states on transport
properties of two-dimensional quantum spin Hall insulators. The bound states
are found able to induce quantum percolation in the originally insulating
bulk.  At some critical antidot densities, the quantum spin Hall phase can
be completely destroyed due to the maximum quantum percolation. For systems
with periodic boundaries, the maximum quantum percolation between the bound
states creates intermediate extended states in the bulk which is originally
gapped and insulating. The antidot induced  bound states plays the same role
as the magnetic field in the quantum Hall effect, both makes electrons go
into cyclotron motions. We also draw an analogy between the quantum
percolation phenomena in this system and that in the network models of
quantum Hall effect.
\end{abstract}

\maketitle

Quantum spin Hall (QSH) insulators have been known as two-dimensional (2D)
topological insulators, which have been studied intensively because of the
novel physics they host.\cite{Kane_RMP, Zhang_RMP,Moore_review} The bulk of
these systems shows insulating gaps but the system boundaries processes
time-reversal symmetry (TRS) protected gapless surface or edge modes, characterized by the $Z_2$ topological invariant. 
\cite{KaneMele_05PRL_1,KaneMele_05PRL_2,BHZ,HgTe_experiment_1,Konig_JPSJ,HgTe_experiment_2}
Disorders are believed not able to cause backscattering of these modes
because of their spin-momentum locking. In two dimension, these
gapless modes are also known as helical edge states that are responsible for
the QSH effect\cite{KaneMele_05PRL_1,KaneMele_05PRL_2,BHZ}. In our previous
works, we have shown that defects in the form of vacancies in 2D topological
insulators can induce bound states that appear within the bulk band gap \cite%
{Shan_defect_BS,Lu_impurity_BS}. Electrons are able to go into cyclotron
motions around the vacancies through these bound states, which is caused by
the non-trivial underlying topology of the system. In reality, defects
inevitably exist in 2D QSH systems. It is however not particularly clear how
these bound states affect the QSH phase as well as the transport properties
of the helical edge states.

In this paper, we study QSH systems in which vacancies appear as arrays
using theoretical models. Such kind of man-made systems have been referred
to as antidot arrays.\cite{FGK_prl,Weiss_prl} In the well-studied 2D
electron systems such as GaAs/GaAs based and InAs based heterojunctions,
antidot arrays display intriguing physical properties.\cite{Bird_book} QSH
insulators are ideal platforms for studying antidot systems since they are
true 2D materials with exotic physical properties. By defining antidot
lattices on the 2D QSH systems, we show that antidots induce extra energy
bands which can greatly smear the band gap. When exceeding the critical
antidot density, the QSH phase can be completely destroyed. When antidots
are randomly distributed, the existence of antidot-induced energy bands and
band gap smearing is confirmed by calculating the density of states of the
2D system. We also study the transport properties in these systems under
different boundary conditions. It is found that quantum percolation occurs
taking the bound states as stepping stones. Physically this happens because
of the finite spatial distribution and overlapping of these bound states. In
the open boundary condition (OBC), the bound states formed transverse
percolation channels allow the backscattering between conducting channels at
opposite edges of the sample with the same spin, which then destroys the QSH
phase (see Fig. 1e). In the closed boundary condition (CBC), the
longitudinal percolation forms new conducting channels between the source and
drain in a two-terminal configuration, as illustrated in Fig. 1d, which can
be viewed as truly extended states. We found that transition from QSH phase
to conventional insulating phase accompanies appearance of these extended
states.
\begin{figure}[h]
\includegraphics[width=0.446 \textwidth]{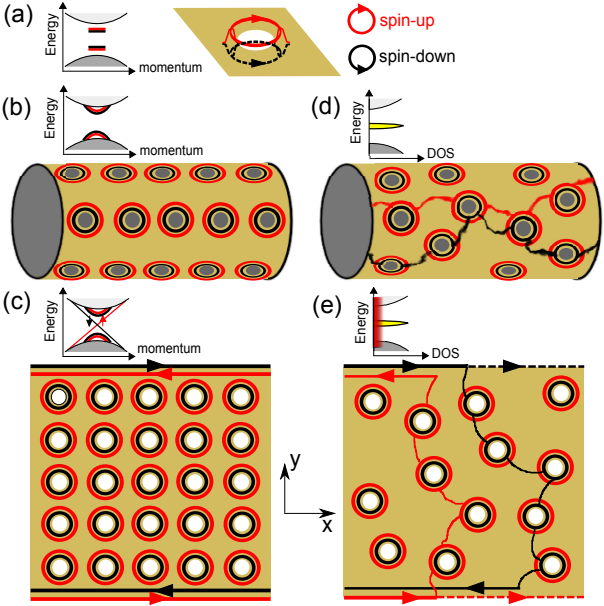}
\caption{(Color online) Schematic description of the influences of in-gap
bound states in 2D TIs. (a) Bound states induced by a single antidot. Red
(black) colors represent the spin-up (-down) component; (b) Antidot lattice
with CBC that is infinite in $x-$direction. Additional antidot induced
energy bands appears in the bulk band gap; (c) Antidot lattice with OBC, the
antidot bands co-exist with the helical edge states; (d) In CBC, at a
critical antidot density, the electron percolates through the overlapping
bound states and form longitudinal conducting channels; (e) In OBC,
transverse percolation channels allows backscattering between edge states on
the opposite edges and finally destroy the QSH phase.}
\label{fig1_Spectrum}
\end{figure}

Following our previous works\cite{Shan_defect_BS,Lu_impurity_BS}, a modified
Dirac model is used to describe the 2D QSH insulators\cite{shen-dirac},
\begin{equation}
H_{0}=v\mathbf{p}\cdot \boldsymbol{\alpha}+\left( mv^{2}-Bp^{2}\right) \beta
,  \label{modified-Dirac}
\end{equation}%
where $-Bp^{2}=-B(p_{x}^{2}+p_{y}^{2})$ is the quadratic correction to the
topological mass term $mv^{2}$, $p_{i}=-i\hbar \partial _{i}$ ($i\in \{x,y\}$%
) is the momentum operator. $v$, $m$ and $B$ have the dimension of speed,
mass and reciprocal mass, respectively. The Dirac matrices satisfy the
anti-commutation relations $\alpha _{i}\alpha _{j}=-\alpha _{j}\alpha _{i}$ (%
$i\neq j$)$,$ $a_{i}\beta =-\beta \alpha _{i}$ and $\alpha _{i}^{2}=\beta
^{2}=1$. A common representation of the Dirac matrices can be expressed as,
\begin{equation}
\alpha _{i}=\sigma _{x}\otimes \sigma _{i},\ \ \beta =\sigma _{z}\otimes
\sigma _{0},  \label{Dirac-matrices}
\end{equation}%
where $\sigma _{i=x,y,z}$ are the Pauli matrices, $\sigma _{0}$ is the $%
2\times 2$ unit matrix, and $\otimes $ represents the Kronecker product. The
Hamiltonian (\ref{modified-Dirac}) preserves time-reversal symmetry $%
\mathcal{T}H_{0}\mathcal{T}^{-1}=H_{0}$, where $\mathcal{T}=-i\alpha
_{x}\alpha _{z}\hat{K}$ is the time-reversal operator and $\hat{K}$ is the
complex conjugate operator. The topological nature of this Hamiltonian has
been discussed in detail in our previous works\cite{shen-dirac}. This system
is topologically non-trivial (with $Z_{2}=1$ ) while $mB>0$, and trivial
(with $Z_{2}=0$ )while $mB<0$.\cite{KaneMele_05PRL_1}

After appropriately reordering the basis, the Hamiltonian (\ref%
{modified-Dirac}) can be decoupled into two blocks\cite%
{Shan_defect_BS,Lu_impurity_BS},
\begin{equation}
h_{\pm }=(mv^{2}-Bp^{2})\sigma _{z}+v(p_{x}\sigma _{x}\pm p_{y}\sigma _{y}),
\label{2D2x2blocks}
\end{equation}%
with $h_{-}$ being the time-reversal counterpart of $h_{+}$. Because the
system keeps TRS, we only consider the $h_{+}$ block. When being discretized
onto an square-lattice, the tight-binding version of $h_{+}$ reads,
\begin{eqnarray}
h_{+}^{\mathrm{TB}} &=&\sum_{\mathbf{i}}C_{\mathbf{i}\uparrow }^{\dag }%
\mathbf{M}C_{\mathbf{i}\uparrow }+\sum_{\mathbf{i}}\left( C_{\mathbf{%
i+\delta _{x}},\uparrow }^{\dag }\mathbf{t_{x}}C_{\mathbf{i}\uparrow }+%
\mathrm{H.c.}\right)   \notag  \label{hplusTB} \\
&&+\sum_{\mathbf{i}}\left( C_{\mathbf{i+\delta _{y}},\uparrow }^{\dag }%
\mathbf{t_{y}}C_{\mathbf{i}\uparrow }+\mathrm{H.c.}\right)
\end{eqnarray}%
where $C_{\mathbf{i}\uparrow }=\left( c_{\mathbf{i}\uparrow },d_{\mathbf{i}%
\uparrow }\right) ^{\mathrm{T}}$ is the annihilation operator of electrons
at site $\mathbf{i}$ with up spins but belonging to two different bands\cite%
{BHZ}, $\mathbf{\delta _{x}}$ $(\mathbf{\delta _{y}})$ is the unit vector
displacement (with length $a$) between two nearest-neighbor sites in the $x$
$(y)$ direction, and the on-site energy and hopping matrices in $x$ $(y)$
directions reads,
\begin{equation}
\mathbf{M}=\left( mv^{2}-\frac{4B\hbar ^{2}}{a^{2}}\right) \sigma _{z},\
\mathbf{t_{x(y)}}=-\frac{i\hbar v}{2a}\sigma _{x(y)}+\frac{B\hbar ^{2}}{a^{2}%
}\sigma _{z}.  \label{2x2Mtxy}
\end{equation}

Numerically, the antidot is modeled by leaving the antidot regions as vacancies on the tight binding
lattice. The CBC geometry is realized by adopting periodic boundary
condition in the $y$ direction (see Fig. 1). Transport features are
investigated through the combination of the non-equilibrium Green's function%
\cite{Datta1} and the Landauer-B$\mathrm{\ddot{u}}$ttiker\cite{LB_formula}
formalism. During the calculation, the effects of each semi-infinite leads
are taken into account by introducing the corresponding self-energy terms.
To clearly illustrate the physics, we adopt such model parameters to
describe the clean substrate: $mv^{2}=-10$, $\hbar v=72$, and $B\hbar
^{2}=-28$, which is in the topologically non-trivial regime, i.e., $mB>0$.
In the tight binding calculations, the lattice constant is set to be $a=1$.

For antidots on a finite size sample, its induced bound state can be
identified by directly diagonalizing the total tight-binding Hamiltonian.
The bound states are states that have spatially localized wave functions and
discrete energy spectrum. The antidot-induced bound states in 2D topological
insulators are found both in the bulk band and in the band gap with wave
function confined around it (Fig. 2a)\cite{Shan_defect_BS,Lu_impurity_BS}.
However, those with energies existing in the bulk band gap are of most
interest to us since they coexist with helical edge states in the QSH
insulator. As an illustration, we plot out the in-gap bound state wave
function for a single antidot and a simple antidot array in Fig.2. As is
shown, the bound states are indeed localized around the antidots. Away from
the center their wave functions quickly decay, which can be approximated in the form $\psi e^{-r/\xi}$.
\cite{Shan_defect_BS,Lu_impurity_BS,shen-dirac}
The characteristic decay length $\xi$ in Fig.2 in theory is about 11a, which agrees well with our numerical results.
For the antidot arrays, as is seen in Fig.2, the
in-gap bound states they induce overlap with each other and eventually
become connected.

\begin{figure}[tbph]
\includegraphics[width=0.47 \textwidth]{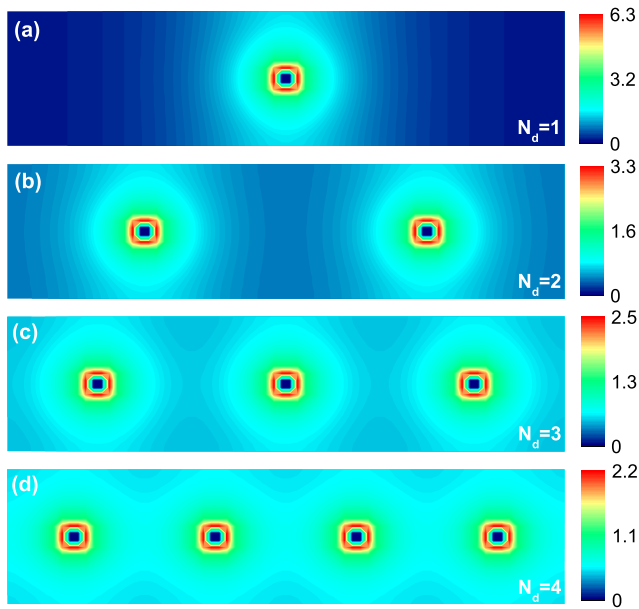}
\caption{(Color online) Connecting of in-gap bound states as the antidots
become dense. $2\times2$ sized antidots are equally spaced on a $20\times72$
torus geometry. (a)-(d) show the evolution of overlapping for in-gap bound
state wave functions with antidot number $N_d=1$ to 4. }
\label{fig2_BS_connecting}
\end{figure}

Consider an antidot array with 100-sites in width. The finite-size
effect-induced energy gap in the clean system of this size is around $%
1.6\times 10^{-3}$,\cite{ZhouBin_PRL_2008}, which is negligible compared
with the bulk gap ($2mv^{2}=20$). The antidots are equally spaced in both $x$
and $y $ directions with the spacing being $D$ sites. We calculate the
spectrum of the antidot lattice with $D=5,4,3,2$ under both open and closed
boundary conditions.  Meanwhile the wave function of the lowest conduction
band at $k_{x}^{0}=\pi /(D+1)$ (red solid circles in Fig. 3) is taken as an
example to demonstrate the emergence of the intermediate states formed by
the bound states in the band gap (Fig. 3 insets).

\begin{figure}[tbph]
\includegraphics[width=0.47 \textwidth]{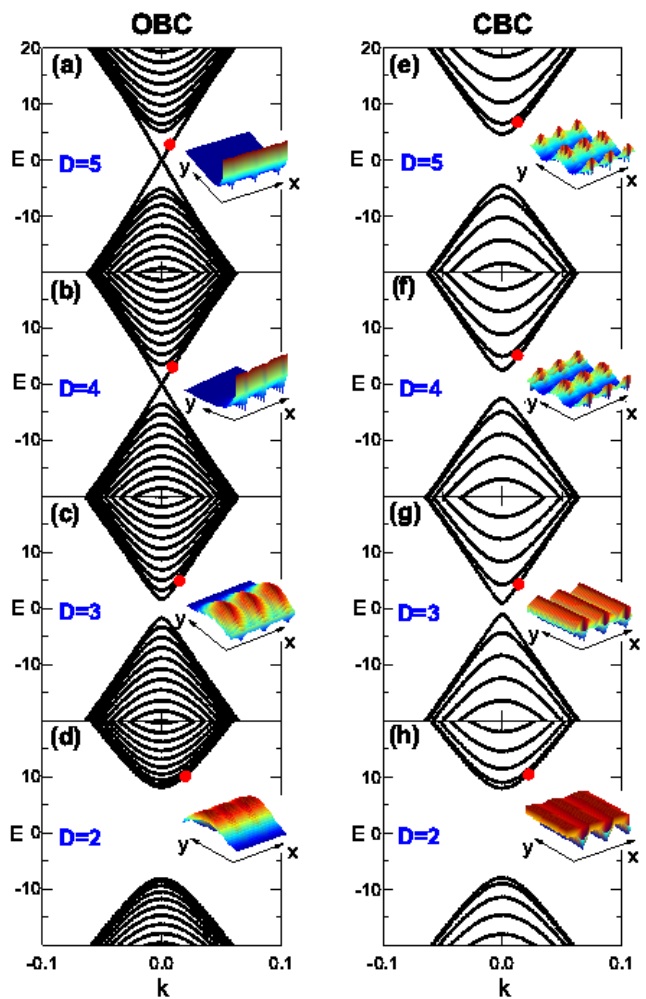}
\caption{(Color online) Energy spectrum of QSH antidot lattices at different
antidot spacing $D$ under OBC and CBC. The width is 100-sites and the
antidot size is $1\times1$ site. (a)-(d) are for OBC, (e)-(h) are for CBC.
The solid red circle in each figure indicate $k_x^0=0.1\frac{\protect\pi}{%
(D+1)}$ on the lowest conduction band, the corresponding wave function for
three principal layers are shown in the insets. }
\label{fig2_spectrum}
\end{figure}

As the antidots become dense, the bulk band gap is found to close up and
then re-opens, as is shown in Fig. 3. When $D=5$ and 4, the edge states
survive under OBC. Meanwhile the antidot-induced energy bands already appear
between $E=\pm 10$ where the bulk band gap was, as is seen in the lowest
conduction band wave functions in CBC in the inset of Fig.3e. When $D$
gradually decreases, the band gap shrinks more. At $D=3$ the edge states in
OBC disappear and the band gap is no longer non-trivial (Fig. 3c). In this
case, the lowest conduction band becomes a bulk mode, as shown in the inset
of Fig. 3(c). Since the antidot size cannot be continuously changed on the
lattice model, this implies that there exists a critical point between $D=4$
and $D=3$ where the band gap vanishes and the QSH phase is destroyed. As the
antidots become denser, the band gap further opens up and no states can be
found in the low energy regime.

After consider the physical picture illustrated in Fig.2, it can be inferred
that there is a critical point where the maximum percolation caused by the
antidot-induced bound states exists. This can be explicitly demonstrated by
calculating the two-terminal transmission coefficient $T_{lr}$ with antidots
randomly distributed in the sample. We choose a 2D sample composed by $%
W\times L=160\times 160$ sites. $1\times 1$ sized antidots are randomly
patterned at a density $p$, which is defined as the number of antidot sites
divided by $WL$. For example, in the antidot array case of Fig.3c and
Fig.3g, the antidot density is $p=1/16$. The Fermi level is fixed at $%
E_{F}^{0}\approx 0$ but carefully avoided from the tiny gap from finite-size
effect. The $T_{lr}$ versus $p$ curves under OBC and CBC are shown in Fig.
4(a) and (b), respectively. For each $p$, the sample averaging is taken with
100 random antidot configurations. Indeed, when $p$ is small, $T_{lr}$
remains one/zero under open/closed boundary condition. But when $p$
approaches a critical value $p_{c}\approx 9/160$, under OBC $T_{lr}$ drops
dramatically while under CBC $T_{lr}$ acquires a sharp peak with the peak
value approaching 1, which is a signature of maximum percolation in the
bulk. When $p\gg p_{c}$ , $T_{lr}$ vanishes under both boundary conditions.
The critical point value we find $p_{c}=9/160$ agrees very well with that we
inferred from the previous results from the antidot arrays in Fig.3. And
thus the physical picture of bound state induced percolation can be
confirmed.

\begin{figure}[tbph]
\includegraphics[width=0.48 \textwidth]{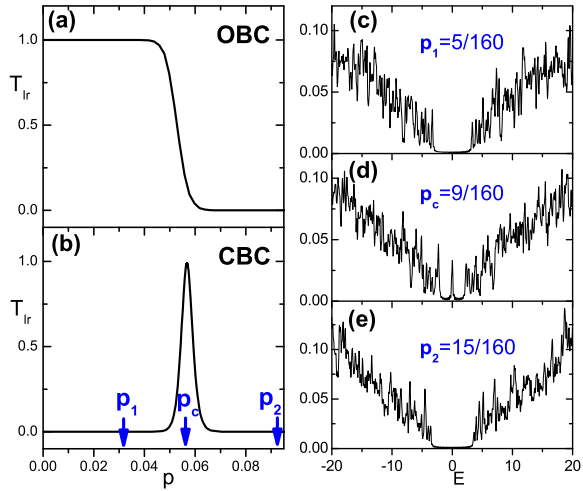}
\caption{(Color online) Quantum percolation of electron through in-gap bound
states induced by randomly distributed antidots. (a) and (b) show the two
terminal transmission coefficient $T_{lr}$ v.s. $p$ curves under OBC and CBC
with sample averaging. The sample is $W\times L=160\times 160$ in size.
(c)-(e) are the corresponding density of states at $p_{1}=\frac{5}{160}$, $%
p_{c}=\frac{9}{160}$ and $p_{2}=\frac{15}{160}$, respectively. In these
calculations, only the $h_{+}$ sub-block is considered. When the
time-reversal symmetry is conserved, the $h_{-}$ block produces the same
result.}
\label{fig3_random}
\end{figure}

The above picture is further confirmed by calculating the system's density
of states. To eliminate the contribution of the edge states in the bulk band
gap, CBC in both $x$ and $y$ directions are adopted. We choose three typical
antidot densities: $p_{1}=5/160$, $p_{c}=9/160$ and $p_{2}=15/160$. For each
density, the averaged density of states is calculated with 100 random
antidot configurations using the kernel polynomial method\cite{KPM}. When $%
p<p_{c}$, the original inverted bulk gap is reduced but still survives to
protect the QSH phase, as is shown in Fig. 4(c). For $p=p_{c}$, a sharp peak
appears at $E=0$ as is shown in Fig. 4(d), which is from the conducting
channel in the bulk formed by the in-gap bound state. The gap seen between
the conduction and valence band in Fig. 4(d) is solely due to the finite
size effect, because it can be inferred from Fig. 3(g) that in the continuum
limit both the valence and conduction band spectrum should become
continuous. When $p>p_{c}$, the band gap re-opens and the system becomes a
conventional insulator [Fig. 4(e)].

It is well established in the quantum Hall effect that the quantum
percolation process is responsible for the transition between the quantized
Hall conductance plateaus\cite{ran_network_model,XiongGang_PRL}, which is a
direct manifestation of the localization-delocalization quantum phase
transition. In the quantum Hall effect, electrons go into cyclotron motions
because of the high perpendicular magnetic field in which process the
disorders help to pin the cyclotron orbitals locally. The quantum
percolation can thus be intuitively understood in the sense that the
cyclotron orbitals connect with each other at some critical external
conditions (e.g.: magnetic field strength, Fermi surface, etc). Based on
this physical picture, the Chalker-Coddington network(CCN) model is
developed which gracefully describes the localization-delocalization
transition in the quantum Hall effect in the presence of disorders\cite%
{ran_network_model,CCN_model}. Upon each of such transition, it is also
accompanied by the change of the topological order of the system, namely the
Thouless-Kohmoto-Nightingale-Nijs integer\cite{TKNN}, which changes by 1 at
each of these transitions.

In the QSH insulator we studied, the maximum percolation accompanied with
destroying of the QSH phase can also be seen as such kind of
localization-delocalization quantum phase transition. But in this case, it
is accompanied by the change in the $Z_{2}$ topological invariant from 1 to
0.\cite{Onoda_prl,TAI} Through our previous papers, it is made clear that the bound states
themselves have helical nature (or chiral when considering only $h_{+}$ or $%
h_{-}$ sub-block)\cite{Shan_defect_BS,Lu_impurity_BS}, which is similar to
the chiral nature of cyclotron orbitals in the quantum Hall effect. By
analogy it is possible that at the critical antidot density, the bound
states connect and form 2D networks in real space, i.e. the formation of
percolation. At this condition, truly extended states exist and the
localization length of electrons should diverge. The percolation provides
new conducting channels in the bulk, which allows edge states be scattered
backward across the sample through the other edge in OBC [as is seen in the
drop of $T_{lr}$ in Fig. 4(a)]. Meanwhile, the existence of extended states
is proven by the dramatic raise of $T_{lr}$ shown in Fig. 4(b). It is noted
that in the QSH insulator, there are always two copies of percolation
channels with opposite chirality, one from $h_{+}$ and the other from $h_{-}$%
.

To conclude, our results indicate that appropriate density of antidots in a
2D QSH insulators can dramatically change the sample's electronic structure
and consequently the transport features. There exists a critical antidot
density at which the bulk electrons becomes truly delocalized because of the
quantum percolation channels across the sample formed by the antidot induced
bound states. This localization-delocalization transition is accompanied by
the vanishing of the QSH phase.

This work was supported by the Research Grant Council of Hong Kong under
Grant No. HKU7051/11P, and by National Natural Science Foundation of China
(Grant No. 11104060).

\end{document}